\documentclass[12pt,preprint]{aastex}
\usepackage[]{natbib,epsfig,graphics,rotating}
\def\msun{M$_\odot$}
\def\msunsp{M$_\odot$ \ }

\def\msunp{M$_\odot$}

\def\cm3{cm$^{-3}$}
\def\kms{km\thinspace s$^{-1}$\ }

\def\13co{$^{13}$CO}
\def\ceio{C$^{18}$O\ }

\slugcomment{To Appear in the Astrophysical Journal}

\shorttitle{Dust Ring in the Coalsack}
\shortauthors{Lada et al.}

\begin{document}


\title{Discovery of a Dusty Ring in the Coalsack:\\
A Dense Core Caught in the Act of Formation?}


\author{Charles J. Lada, and Tracy L. Huard}
\affil{Harvard-Smithsonian Center for Astrophysics, 60 Garden Street,
Cambridge MA 02138}
\email{clada@cfa.harvard.edu thuard@cfa.harvard.edu}

\author{Lionel J. Crews}
\affil{Department of Geology, Geography, and Physics, University of 
Tennessee at Martin, Martin, TN 38238}
\email{lcrews@utm.edu}

\and

\author{Jo\~ao F. Alves}
\affil{European Southern Observatory, Garching Germany}
\email{jalves@eso.org}

\vfill\eject

\begin{abstract}

We present a new infrared extinction study of Globule 2, the most opaque
molecular cloud core in the Coalsack complex.  Using deep
near-infrared imaging observations obtained with the ESO NTT we are able to examine the
structure of the globule in significantly greater detail than previously possible.  We
find the most prominent structural feature of this globule to be a strong central ring
of dust column density which was not evident in lower resolution studies of this cloud.
%%For a spherical cloud, this ring would represent a shell of high density and
%%pressure that would not be in dynamical equilibrium with its surroundings.  
%This ring represents a region of high density and pressure that is probably not 
%in dynamical equilibrium with its surroundings. In
%particular, there appear to be no sources of additional pressure in the central regions
%of the cloud to 
%%support the dense shell against gravity and 
%prevent its inward
%implosion.  
This ring represents a region of high density and pressure that is likely a
transient structure.  For a spherical cloud
geometry the ring would correspond to a dense inner shell of high pressure that
could not be in dynamical equilibrium with its surroundings since there appear
to be no sources of pressure in the central regions of the cloud that
could support the shell against gravity and prevent its inward implosion.
The timescale for the inward collapse of the ring would be less than 2
x 10$^5$ years, suggesting that this globule is in an extremely early stage of
evolution, and perhaps caught in the process of forming a centrally condensed dense core
or Bok globule.  Outside its central regions the globule displays a well-behaved density
profile whose shape is very similar to that of a stable Bonnor-Ebert sphere.  Using SEST we
also obtained a \ceio spectrum toward the center of the cloud.  The CO observation
indicates that the globule is a gravitationally bound object.  Analysis of the CO line
profile reveals significant non-thermal gas motions likely due to turbulence.  As a
whole the globule may be evolving to a global state of quasi-static dynamical 
equilibrium in which thermal and turbulent pressure balance gravity.

\end{abstract}



\section{Introduction}

Low mass dense cores found within molecular cloud complexes and isolated dark globules
(also known as Bok globules) are the simplest configurations of dense molecular gas and
dust known to form stars (e.g, Benson \& Myers 1989; Yun \& Clemens 1990).  They have
been long recognized as important laboratories for investigating the physical processes
which lead to the formation of stars and planets (e.g., Bok 1948).  In recent years deep
infrared imaging surveys of such globules have provided an important new tool for
detailed examination of their structure.  For example, infrared observations of the
starless globule B 68 produced an exquisitely detailed measurement of the cloud's radial
density distribution (Alves, Lada \& Lada 2001, hereafter ALL01).  This density
distribution was shown to match very closely that predicted for a marginally stable,
pressure bounded, isothermal sphere in hydrostatic equilibrium (i.e., a Bonnor-Ebert
sphere).  Similar deep near-infrared observations of the darker globule B 335 enabled a
detailed investigation of the radial density profile of a globule containing a central
protostar.  The profile in this star forming globule was found to be significantly
steeper than that of B 68, consistent with expectations for a collapsed or collapsing
core (Harvey et al. 2001).  These observations suggest that the evolution of dense cores
to form stars can be effectively traced and measured by detailed observations of the
cloud's density structure, best obtained by infrared extinction observations.

Two of the least understood aspects of the star formation process are the initial
conditions that describe dense cores that ultimately form stars and the origin of such
dense cores from more diffuse atomic and molecular material.  Deep infrared and
millimeter-wave observations of starless cores or globules offer the best opportunity to
investigate these issues.  Starless cores account for about 70\% of optically selected
dark cores and globules (Lee \& Myers 1999).  A particularly interesting group of cores
in this regard are those associated with the conspicuous Coalsack dark cloud complex in
the southern Milky Way (Tapia 1973; Bok, Sim \& Hawarden 1977; Bok 1977).  The entire
Coalsack complex subtends an angle of roughly 6$^o$ on the sky corresponding to a linear
dimension of nearly 15 pc at the distance of 150 pc estimated for this cloud (e.g.,
Cambresy 1999, and references therein).  A survey of $^{12}$CO emission found the cloud
to be characterized by complex, filamentary structure and to be relatively massive
containing about 3500 \msunsp of material (Nyman, Bronfman \& Thaddeus 1989).  However,
a survey of emission from the rarer $^{13}$CO isotopic line found the ratio of $^{13}$CO
to $^{12}$CO emitting gas in the cloud to be considerably lower (17\%) than that
(50-80\%) which characterizes nearby star forming molecular clouds (Kato et al.  1999).
This observation indicated that the fraction of dense gas in the Coalsack complex is
considerably smaller than that which characterizes typical star forming cloud complexes.
The paucity of dense gas coupled with the lack of the usual signposts of star formation
activity, such as emission-line stars, HH objects, embedded infrared sources, etc.
(e.g., Kato et al.  1999) suggests that the Coalsack may be a molecular cloud
complex in the earliest stages of evolution.  The globules within the Coalsack are
apparently all starless and also may be in the early phases of development.

Knowledge of the detailed structure of globules in the Coalsack may provide further
insight concerning their physical nature and evolutionary status.  In particular, it
would be interesting to know to what extent the structure of a Coalsack globule
resembles that of a starless cloud like B 68 or a star forming globule such as B 335.
Tapia's Globule~2 is the most prominent and likely densest globule in the Coalsack (Bok
et al.  1977; Kato et al.  1999).  Located slightly below ($\sim$ 1$^o$) the galactic
plane, it is projected against a rich star field and is a prime candidate for infrared
extinction studies.  Jones et al.  (1980) obtained JHK photometry of 75 stars located
behind the Coalsack cloud and derived individual extinctions to all these stars.  Their
observations suggested a radial column density distribution that was not highly
centrally concentrated.  Racca, Gomez and Kenyon (2002) obtained JHK infrared images of
the globule that were more sensitive and resulted in the detection of a few thousand
stars behind the cloud.  They used H-band star counts to construct a low resolution
extinction map of the globule and derived an azimuthally averaged radial density profile
which confirmed the shallow nature of the radial column density distribution.  Moreover,
they found that the radial density distribution could be fit by a Bonnor-Ebert
configuration with a density profile quite similar to that of B 68 and considerably
shallower than that of B 335.

In this paper we report new and deeper near-infrared imaging observations of Globule~2
(hereafter G2).  We use these observations to determine the individual line-of-sight
extinctions to thousands of stars behind the cloud.  These measurements enable an
examination of the structure of the globule to be made in a degree of detail
considerably greater than that obtained by previous studies and comparable to that
acheived earlier for B 68.  The high angular resolution achieved by our observations
reveal the globule to be more structured than suspected previously.  Moreover, these
observations indicate that the spatial distribution of column density in this globule
differs significantly from that of other well studied examples, such as B 68 and B 335.
In particular, the radial dust column density profile of G2 is not only found to be
shallow, but also to be characterized by a significant central depression that, in turn,
is surrounded by a prominent ring of high column density.  These characteristics suggest
that Globule~2 may represent a very early stage in the evolution of cloudy material to
form a centrally condensed dense core.

\section{Telescopes and Instrumentation}

We used the ESO 3.5-m New Technology Telescope (NTT) at the 
European Southern Observatory (ESO) in La Silla, Chile on
8-9 March 1999, 13 March 2000 and 18 June 2003
to obtain the imaging observations reported here. The telescope
was outfitted with the infrared imager/spectrometer SOFI (Moorwood
et al. 1998). SOFI is equipped with a 1024 $\times$ 1024 pixel Rockwell
array which was configured to provide a field of view $\sim$ 5 $\times$ 5
arc min with a spatial resolution of 0.292 $\pm$ 0.001 arc sec pixel$^{-1}$.
Observations were obtained with J (1.25 $\mu$m), H (1.65 $\mu$m), and
K$_s$ (2.16 $\mu$m) filters.

We also used the 15-m Swedish ESO Submillimeter Telescope (SEST)
at La Silla on 28 May 2000 to obtain an observation of the C$^{18}$O
(J = 2--1) line at 219 GHz toward the center of the globule. The
receiver was equipped with an SIS mixer which produced a system
temperature of approximately 200 K. The half-power beamwidth at
the observing frequency was approximately 24 arc sec. An AOS
spectrometer provided a velocity resolution of 0.06 km sec$^{-1}$.
A series of five minute, frequency-switched observations were
summed to achieve a total integration time of 60 minutes. This
resulted in an rms noise level of about 0.02 K. 
The calibration was done using the standard chopper wheel method
which compared sky emission to an ambient temperature load to set
the temperature scale for the spectra.

\section{Observations, Results and Analysis}

Figure~\ref{optical} shows an optical image of the region around Globule 
2 in the Coalsack. 
In order to fully image the globule, we obtained a 3 $\times$ 3 mosaic of
J, H and K$_s$ images covering a field approximately 14 $\times$ 14 arc min
approximately centered on the globule. The surveyed region is indicated
by the box in Figure~\ref{optical}.
The total effective exposure times were 50, 30, and 20 seconds for the J, H
and K$_s$ bands respectively, except in the central 5 $\times$ 5 arc min
region where we observed for 95 seconds at J. Each integration consisted of a series
of coadded short dithered observations. The dithers were randomly generated
within a box 40 $\times$ 40 arc sec in size centered on each
field in the mosaic.
The data reduction was accomplished with a combination of standard
Image Reduction and Analysis Facility (IRAF)\footnote{IRAF is distributed
by the National Optical Astronomy Observatory, which is operated by
AURA under contract to the NSF.} routines and custom
software.
Sources were identified and automatically 
extracted from the images using the SExtractor program. All images were 
inspected to find and extract any additional 
stars that were not automatically identified. Aperture
photometry for all sources was obtained using standard
{\it{daophot}} packages in IRAF, while their positions were derived
by comparison with known positions of stars, as listed in the USNO-A2.0
catalog, that were observed in our fields.  Instrumental magnitudes
obtained from  2.3 arc sec diameter synthetic apertures centered
on these sources, were calibrated using the J, H and K magnitudes of stars
within our fields that were listed in the 2MASS catalog. Only stars having
2MASS H-K colors between 0 and 0.3 magnitudes were used for calibration to
avoid possible color corrections between the SOFI and 2MASS photometry for more
reddened stars.
More than 24,000 sources were identified, the vast majority of which 
were detected in both the H and K bands. We estimate our completeness limits
to be about 17th magnitude at H and K and 18th magnitude at J.

\subsection{Extinction Map}

We used the NICE method (Lada et al.  1994, Alves et al.  1998) to derive
extinctions to approximately 24,000 individual stars detected in our deep infrared
survey of Globule 2 and to construct a detailed, high resolution map of
the distribution of extinction across the cloud.  Briefly, we determine the
color excess, $E(H-K)$, for each star:

\begin{equation}
E(H-K) = (H-K)_{obs} - (H-K)_{intrinsic}
\end{equation}

\noindent 
where $(H-K)_{obs}$ is the observed color of the star and $(H-K)_{intrinsic}$ is
the intrinsic color of the star.  The intrinsic $(H-K)$ color for each star was
assumed to be the same as the average color of stars in a nearby, unreddened
control field.  Using data from the 2MASS survey we derived $(H-K)_{intrinsic}$ to
be 0.201 $\pm$ 0.004 magnitudes.
%The 2MASS survey was used for this purpose since none of our NTT
%fields were found to be free of extinction.    
Following convention we scaled the color excesses to equivalent visual extinctions,
A$_V$s, using a standard reddening law, i.e., A$_V$ = 15.9 $E(H-K)$.

\begin{figure}
\plotone{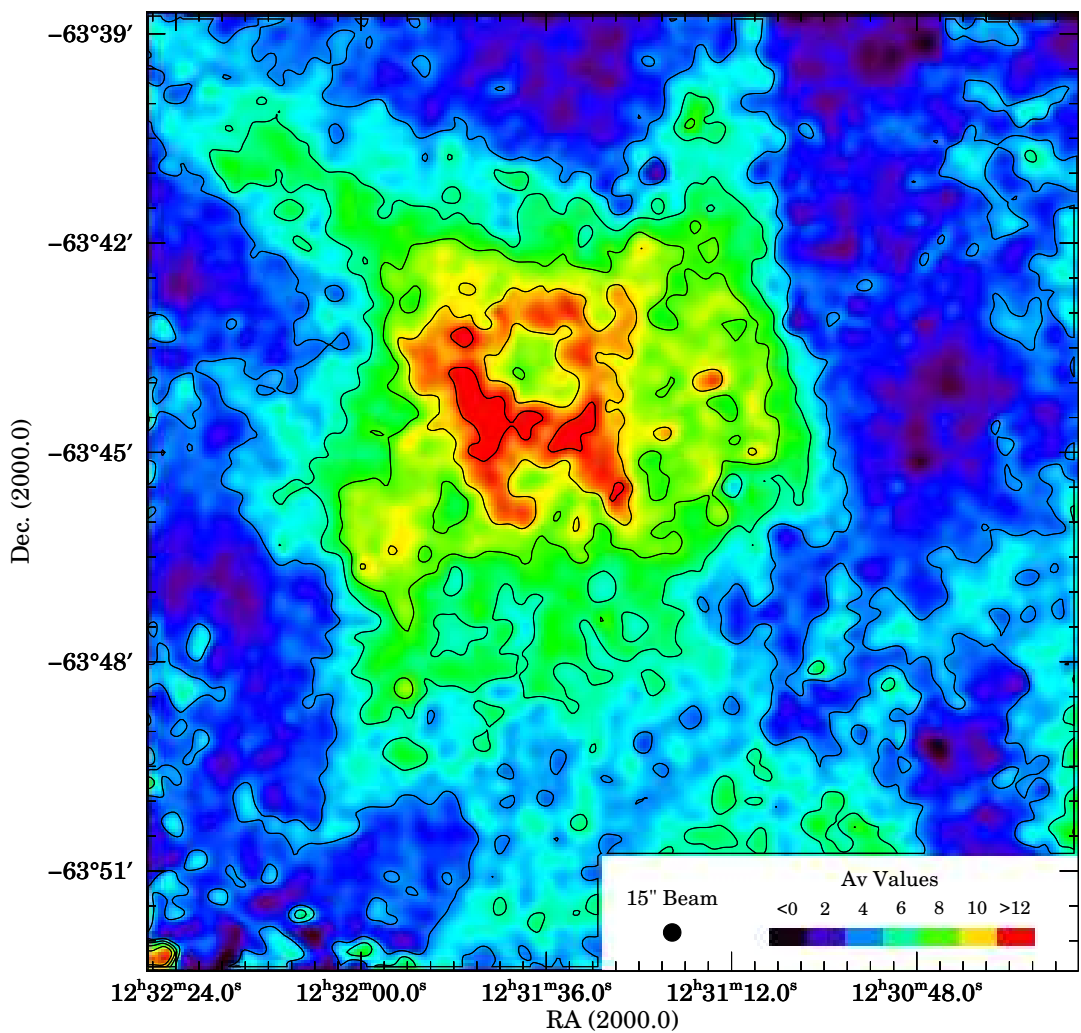}
%\plotone{f2_CMYK.eps}
\vskip -1.0in
\caption{Map of the equivalent visual extinction, A$_V$, in Globule 2 of
the Coalsack cloud derived from deep infrared imaging observations. The
map was constructed by convolving a 15 arc sec gaussian smoothing function
with the pencil-beam extinction measurements of approximately 24,000 stars 
randomly distributed across the mapped region. A 
prominent ring of high column density dominates the structure of this
cloud. Contours (thin lines) begin at an A$_V$ of 4 magnitudes and increase 
in steps of 2 magnitudes. 
\label{extmap}}
\end{figure}

In Figure~\ref{extmap} we present the extinction map for Globule 2 that we
derived from our observations by spatially convolving our data 
with a 15 arc sec gaussian function and sampling at the Nyquist frequency. 
Figure~\ref{extmap} represents the most detailed map of the globule's extinction
yet obtained.  It is a significant improvement over the star count map of Racca et
al.  (1999) primarily because the NICE method produces a direct measurement of the
extinctions to individual stars in each pixel rather than a simple count of the
numbers of stars.  This produces both a more precise and accurate measurement of
extinction in each pixel and enables higher angular resolution to be achieved
because the uncertainties are less sensitive to counting statistics.  Moreover,
our deep survey detected about four times as many stars as in the Racca et al.
map.  The studies of Jones et al.  1980 and Jones et al.  (1984) also used
measurements of infrared color excess ( $E(H-K)$ ) to individual stars derive the
distribution of extinction toward this globule, but they were able to obtain
measurements to only about 100 stars behind the globule, more than two orders of
magnitude fewer stars than observed in this survey.

Our map shows the globule to exhibit considerable structure, the most prominent
feature being a well defined ring at the center of the globule.  A hint of the
ring structure is also evident in the lower resolution star count maps of Racca et
al.  The equivalent visual extinction, A$_V$, produced by the globule ranges from
about 3.5 magnitudes in the outer regions to approximately 12 magnitudes in the
ring.  The map also shows the presence of whispy or filamentary-like structure
outside the high A$_V$ ring.  At this level of spatial resolution, the cloud does
not appear to be as smooth as the B68 cloud which was mapped at similarly high
sensitivity and angular resolution (Alves et al. 2001, 2004).

By spatially integrating the extinction map over the area of the globule we derive the
total mass of the globule to be 15.1 $\pm$ 0.2 \msunsp for the assumed distance of 150 pc
which reasonably agrees with the earlier estimates (11 \msun) of Jones et al.  (1980).  The
quoted uncertainty is primarily due to the  uncertainty in the intrinsic colors of
the background stars.  If we correct for the foreground/background extinction of the more
diffuse Coalsack complex (A$_V$ $\approx$ 3.5 mag) we estimate a lower mass for the globule
of 6.1 $\pm$ 0.5 \msunsp, where the uncertainty is entirely systematic and dominated by the
uncertainty in the foreground/background extinction.  The Jean's mass of the cloud can be
derived from its temperature and mean density, that is, M$_{J}$ = 1.7 $T^{3 \over 2}$
\=n$^{-{1\over 2}}$ \msunp.  For $T$ = 10 K and \=n = 2.7 x 10$^3$ \cm3 we find M$_J$ =
10.4 \msun.  The globule's mass is less than a Jean's mass, suggesting that the globule is
globally stable against collapse.

\subsection{Radial Density Distribution}

Figure~\ref{rprofile} displays the radial profile of extinction that we derived for
G2.  This profile was constructed by azimuthally averaging extinction
measurments for stars within circular annuli, 20 arc seconds wide, sampled at
the Nyquist frequency and centered at the extinction minimum near the geometric
center of symmetry for the map (i.e., $\alpha$ = 12$^{h}$ 31$^m$ 38.6$^s$ and
$\delta$ = -63$^o$ 43' 42.5").  The radial profile of extinction displays a clear
peak of A$_v$ $\sim$ 11.5 magnitudes at a radial distance of 55 arc sec.  The
off-center peak in the radial density profile corresponds to the ring structure
observed in the contour map of figure~\ref{extmap}.  At large radii the extinction
profile smoothly declines until it reachs a plateau at A$_v$ $\sim$ 3.5 magnitudes
near r $\sim$ 290 arc sec.  Noted also by Jones et al.  (1984) and Racca et al.
(1999), this plateau is likely the result of the fact that G2 is embedded in
the Coalsack cloud complex.  However, we note that inspection of Figure~
\ref{extmap} indicates that the extinction due to the general background of
material in the Coalsack is perhaps more patchy than suggested by the constant
plateau in Figure~\ref{rprofile}.

For comparison we plot in Figure~\ref{rprofile} the predicted column density profile
for a pressure truncated isothermal sphere, the so-called Bonnor-Ebert (BE) sphere.
The BE profile was fitted in a least-squares iterative procedure to only those points
beyond a radius of 60 arc sec, that is, from the center of the ring outward.  In
addition a constant plateau of background extinction was included in the BE models to
correspond to the observations.  The BE density profile is clearly much more centrally
concentrated than that of G2.  However the shape of G2's density profile at increasing
radii just outside the ring is well matched by the BE curve.  Even though the density
profile of G2 departs from that expected for a BE configuration in its central
regions, a more detailed comparison of the two profiles can nonetheless provide some
useful insights about the physical nature of the cloud.

A BE sphere is a pressure-truncated isothermal ball of gas within which internal
pressure everywhere precisely balances the inward push of self-gravity and external
surface pressure.  The fluid equation that describes such a self-gravitating,
isothermal sphere in hydrostatic equilibrium is the following well known variant of
the Lane-Emden equation:

\begin{equation}
{1 \over \xi^2}{d \over d\xi}(\xi^2{d\psi \over d\xi}) = e^{-\psi}
\end{equation}

\noindent
where $\xi$ is the dimensionless radius:

\begin{equation}
\xi = {\rm r}/r_c
\end{equation}

\noindent
and $r_c$, is the characteristic or scale radius, 

\begin{equation}
r_c = c_s / (4 \pi G \rho_0)^{1/2}, 
\end{equation}

\noindent
where $c_s$ is the sound speed in the cloud and $\rho_0$ is the density 
at the origin. Equation 2 is Possion's equation in dimensionless form
where $\psi({\xi})$ is the dimensionless potential and is set by the
requirement of hydrostatic equilibrium to be $\psi(\xi)$ = -ln($\rho / \rho_0$). 
The equation can be solved using the boundary conditions that
the function $\psi$ and its first derivative are zero at the origin.
Equation 2 has an infinite family of solutions that are characterized 
by a single parameter, the dimensionless radius at outer edge (R) of the sphere:

\begin{equation}
 \xi_{max} = {\rm R}/r_c.  
\end{equation}

\noindent 
Each solution thus corresponds to a truncation of the infinite isothermal
sphere at a different outer radius, R.  
%The BE density profile is that particular
%density gradient which enables the precise balance between internal pressure
%forces, gravity and external pressure.  The shape of the BE density profile for a
%cloud of fixed size R is identical to that of an infinite isothermal sphere in
%hydrostatic equilibrium out to the same radius R.  
The external pressure at a given
R must then be equal to that which would be produced by the weight of material that
otherwise would extend from R to infinity in an infinite isothermal sphere.  The
shape of the BE density profile for a pressure truncated isothermal cloud therefore
depends on the single parameter $\xi_{max}$.  As it turns out, the higher the value
of $\xi_{max}$ the more centrally concentrated the cloud is.  The stability of such
pressure truncated clouds was investigated by Bonnor (1956) and Ebert (1955)
who showed that when $\xi_{max}$ $>$ 6.5 the clouds are in a state of  unstable
equilibrium, susceptible to gravitational collapse.

Because G2 is embedded in the more extended Coalsack cloud complex we can
only trace the extent of the globule's column density profile to the point where the
extinction is about 3.5 magnitudes, the level of the general background extinction in
the Coalsack complex.  This occurs at an angular radial distance of roughly 290
arc sec.  This gives us only a lower limit to the actual radius of the globule.  For
an angular radius of $\Theta_{max}$ = 290 arc sec, $\xi_{max}$ = 5.8 from the BE
profile fit.  This value of $\xi_{max}$ does not represent a particularly high degree
of central concentration and is consistent with a relatively stable configuration.  For
comparison, the measured extinction profiles for the B 68 and B 335 clouds yield
values of 6.9 and 12.5 for $\xi_{max}$, respectively (ALL01, Harvey et al.  2001),
indicating higher degrees of central concentration.  Apparently the starless B 68
cloud is near the critical state.  The high value of $\xi_{max}$ derived for B 335
suggests a very unstable configuration for that object which is consistent with the
fact that a well developed protostar has already formed in its center.  Of course, the
value of $\xi_{max}$ derived here for G2 must be considered a lower limit to the true
value of $\xi_{max}$ for the cloud since the background extinction due to the Coalsack
complex prevents us from determining the actual outer edge of the globule.  However,
it is likely that the true outer radius of the globule is not much larger than
estimated here.  Consider, for example, that if $\xi_{max}$=6.5, the predicted surface
pressure for the globule would be P$_s$/k = 2 x 10$^4$ \cm3 K.  The external pressure
at the cloud surface, 
P$_{ext}$, provided by the inter-core gas of the Coalsack cloud can be estimated to be
P$_{ext}$ = $\rho v^2$ where $\rho$ is the density of the inter-core gas and $v$ its
turbulent velocity.  For for an interclump density of 10$^2$ cm$^{-3}$, characteristic
of CO emitting regions in molecular clouds, and a CO linewidth of 1.2 \kms which is
observed for the $^{12}$CO gas in the general complex (Nyman et al 1988), we find
P$_{ext}$/k = 2 x 10$^4$ \cm3 K, which is comparable to the surface pressure of 
the BE cloud at a dimensionless radius of $\xi_{max}$ = 6.5.  For larger
values of $\xi$ the density and surface pressure will be lower and not in
equilibrium with the external pressure.  Thus, it appears likely that the cloud does
not extend significantly beyond an angular radius of 290 arc sec and consequently is 
likely to be in a sub- or near- critical state and relatively stable. Indeed, the 
fact that the mass of the globule is found to be sub-Jeans is also consistent with
$\xi_{max}$ being less than 6.5.

Our fit
does indicate that the density profile of this cloud between angular radii of
approximately 50 - 300 arc sec is the same shape as that of a BE sphere out to a
dimensional radius of $\xi$ = 5.8. This enables us to scale the physical
parameters of the cloud using the above equations for $\xi_{max}$ and 
assuming a normal gas-to-dust ratio.  Specifically,

\begin{equation}
\xi_{max} = 10^{-8} \sqrt{{{\rm R} A_v \over \kappa(\xi_{max}) T}}
\end{equation}

\noindent
where $A_v$ is the extinction through the center of the BE cloud, $T$ is the
temperature characterizing the gas, and $\kappa(\xi_{max})$ is the dimensionless
column density resulting from the integration of the dimensionless radial density
profile through the BE cloud. 
That is:

\begin{equation}
\kappa(\xi_{max}) = \int_0^1 {\rho(r) \over \rho_0} d({\xi \over \xi_{max}})
\end{equation}

Once the shape of the density profile ($\xi_{max}$) is fixed, knowledge of any two of the
remaining three physical parameters ($A_v$, R, or $T$) fixes the third.  Thus for
$\xi_{max}$ = 5.8, ($\kappa(5.8)$ = 0.91), $A_v$ = 9 magnitudes (set by the BE fit to the
data), $\Theta_{max}$ = 290 arc sec and a distance of 150 pc, we derive a temperature for
the cloud of $T_{BE}$ = 19 K.  The central density, n$_0$ of the Bonnor-Ebert model is 1.8
x 10$^4$ cm$^{-3}$.  The derived BE temperature is higher than the kinetic temperature of
10 K suggested by CO observations of the Coalsack complex (e.g., Nyman et al 1989).
Reducing the distance to the cloud to 100 pc would reduce the BE derived temperature close
to 10 K.  However, as discussed below, CO observations also indicate that the velocity
field of the globule is characterized by significant non-thermal motions and the derived
BE temperature need not be reflective of the kinetic temperature in the gas.  Instead, the
higher derived temperature may indicate the importance of additional sources of pressure,
such as turbulence or magnetic fields, for the internal support of the cloud (e.g., Lai et
al.  2003).

In the final analysis however, the fact that the observed extinction profile clearly
departs from the BE profile in the central regions suggests that physical parameters
derived from the fitted BE profile must be regarded with appropriate caution.  However, it
is interesting to note here that integrating under the two column density profiles results 
in a comparable total mass ($\sim$ 6 \msun) for the cloud.  The mass
deficit corresponding to the difference between the BE and observed density profiles in the
center of the cloud is approximately 0.1 \msun \ or only roughly 1-2 \% of the total cloud
mass.  In other words, a small redistribution of mass to the center of the cloud could
result in a configuration with a central density concentration similar to that of a BE
sphere or a more typical cloud core.  Thus, although this cloud is not strictly a BE
configuration, it still may not be very far from a state of overall global equilibrium.

\subsection{The C$^{18}$O Molecular-Line Profile}

\begin{figure}

%\centerline{\epsfig{figure=c18o_coalsack_expand.EPS,width=6.5in}}
\centerline{\epsfig{figure=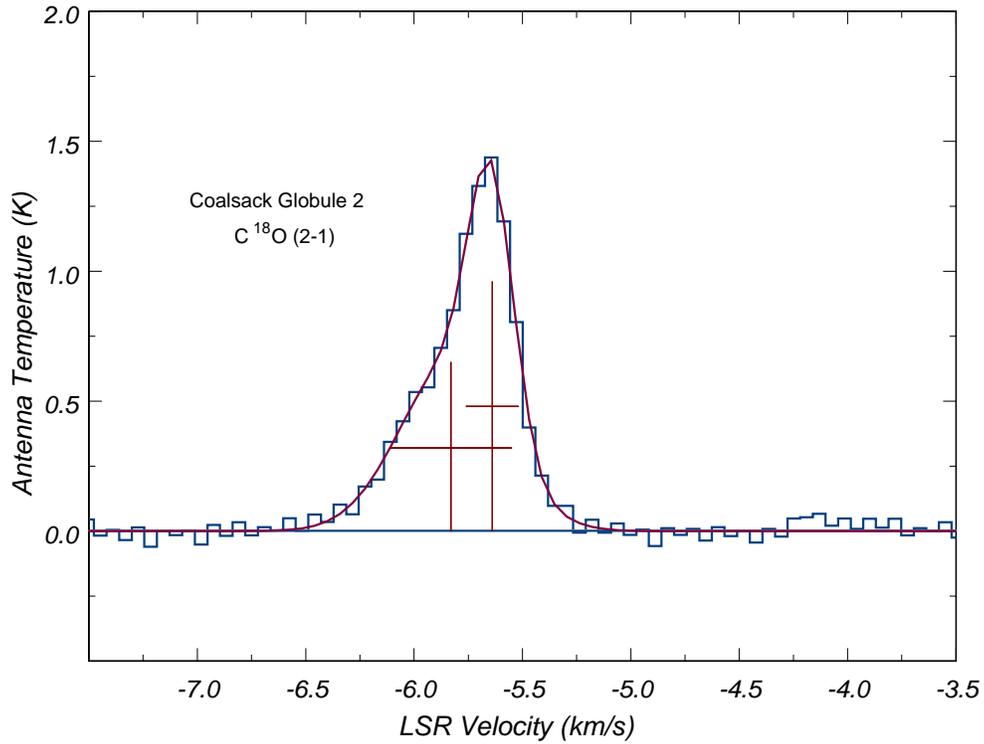,width=6.5in}}
%\plotone{f4.eps}
%\vskip -1.0in
\caption{The observed line profile of the J = 2 $\rightarrow$ 1 transition 
of \ceio emission toward the central regions of the globule. The profile
consists of a blend of two components whose parameters were derived from a 
simultaneous fit of two gaussian functions and a linear baseline (smooth line)
to the observed spectrum. The crosses mark the locations, amplitudes and
linewidths (FWHP) for the two components derived from the gaussian fitting. 
\label{coprofile}}
\end{figure}

Figure~\ref{coprofile} shows the profile of the C$^{18}$O 2--1 emission line that we
obtained near the center of the globule at the coordinates of $\alpha$(2000) = 12$^h$
31$^m$ 34$^s$ and $\delta$(2000) = -63$^o$ 44' 51''.  The line appears to be a blend of
two components.  A simultaneous fit of two gaussian functions to the observed profile
reveals the profile to consist of a relatively bright narrow-line component (NLC)
blended with a fainter broad-line component (BLC).  Specifically, the derived parameters
for the peak line temperatures, LSR center velocities and linewidths for the two
components of the profile are:  0.96 K $\pm$ 0.03 K, -5.64 $\pm$ 0.02 km s$^{-1}$, 0.25
$\pm$ 0.02 km s$^{-1}$ and 0.65 $\pm$ 0.03 K, -5.83 $\pm$ 0.02 \kms, and 0.55 $\pm$ 0.02
\kms, respectively.  The narrow component is redshifted with respect to the broad
component with the relative line-of-sight velocity difference being comparable to the
sound speed (0.20 \kms) in a 10 K molecular hydrogen gas.  The observed linewidths of
the CO emission features in G2 are somewhat smaller than those ($\sim$ 0.7 \kms) which
often characterize globules and cloud cores (e.g., Tachihara, et al.  2002).  On the
other hand, the G2 linewidths are not as narrow as those in the B 68 cloud whose \ceio
emission is characterized by a single, thermally broadened component (Lada et al.
2003).

 The blended, dual component nature of the molecular emission line profile was not
evident or reported in previous $^{12}$CO (Nyman, Bronfman, \& Thaddeus 1989), \ceio
(Kato et al.  1999), NH$_3$ emission (Bourke et al.  1995) or H$_2$CO absorption
(Brooks, Sinclair \& Manefield, 1976) observations of this source.  However, most of
these earlier observations were made with considerably lower angular resolution than
that (24 arc sec) which characterizes our observation.  For example, strong H$_2$CO
absorption obtained with an angular resolution of 4.4 arc min by Brooks et al (1976)
appears consistent with a single line characterized by a linewidth (corrected for
instrumental broadening) of 0.6 \kms and a line center velocity of V$_{lsr}$ = -5.8
\kms, which is close to the parameters of the BLC of our \ceio profile.
Since the beamwidth of the H$_2$CO observations is a factor of ten larger than that of
our 2-1 \ceio observations, the lack of evidence for a second, narrow component in the
H$_2$CO line profile suggests that the narrow component present in our \ceio spectrum
is not as spatially extended as the broad component.

Both components are broader than would be expected (0.12 \kms) for \ceio molecules in a
gas characterized by a temperature of 10 K, indicating a significant non-thermal
component to the linewidth.  The non-thermal component of the linewidth can be
characterized by a temperature, that is $T_{NT} = (\Delta V_{NT})^2\mu / (8ln2)k $,
where $\mu$ is the mean molecular mass (2.34 m$_H$) and $\Delta V_{NT}$ is the
non-thermal component of the linewidth.  These "non-thermal" temperatures are found to
be 14 K and 2 K for the broad and narrow components respectively.  In the situation
where the total pressure in the cloud is a combination of thermal and turbulent
pressure components, we can write P$_{tot}$ $=$ nk(T$_{eff}$) where T$_{eff}$ = T$_K$ +
T$_{NT}$.  Considering each line component separately we find T$_{eff}$ $=$ 24 K and 12
K, respectively.  The BE temperature of 19 K derived earlier lies between these two
values and its value is thus consistent with the notion that the cloud may not be far
from a state of quasi-static dynamical equilibrium in which gravity is balanced by
thermal plus turbulent pressure (provided that $\Delta V_{NT} \not= \Delta V_{NT}(r)$).
The three dimensional non-thermal velocity dispersions for the two different lines are
found to be 0.39 and 0.15 \kms for the broad and narrow component, respectively.  The
BLC is characterized by supersonic motions while the NLC by subsonic motions.  These
motions are not large enough to unbind the cloud even in the absence of any external
pressure.  We estimate the escape velocity from this cloud ($\sqrt{2GM/R}$) to be
approximately 0.5 \kms which is larger than the velocity dispersion of the broad-line
gas, while the estimated virial velocity ($\sqrt{GM/R}$) of 0.4 \kms is comparable to
the velocity dispersion of the broad-line gas.  Thus the globule is most likely a bound
object.

\subsection{The Sigma-A$_V$ Relation: Small Scale Structure}

Both the extinction map and radial profile derived from the data represent
ordered and uniform spatial samplings of the observations which reveal
structural information about the cloud down to the angular scale of the
sampling function used to construct them.  Structural information on smaller
angular scales is present in the data but lost in the construction of the
ordered extinction map and radial profile.  The data used to produce these maps
consists of more than 24,000 precise, individual extinction measurements.
These measurements are randomly distributed across the cloud with relatively
poor spatial sampling.  However they are characterized by exquisitely high,
pencil-beam, angular resolution.  Various techniques can be exploited to recover
information on spatial scales smaller than those used in making the ordered
extinction map and radial profile.  One technique is to analyze the relation
between $\sigma_{A_V}$, the dispersion in a square pixel of an extinction map
and corresponding mean A$_V$ derived for that pixel.  Lada et al.  (1994)
showed that useful information about the structure of the cloud on angular
scales smaller than the pixel size of the map can be ascertained from the
$\sigma_{\rm A_V}$ -- A$_V$ relation.  In particular, this relation can be used to
set interesting constraints on the amplitude of small scale random spatial
fluctuations in a cloud (Lada et al.  1999, Alves et al.  2004), provided that
any systematic column density gradients are resolved by the
observations.  The amplitude of such fluctuations may be a useful metric to
describe the degree of supersonic turbulence in a cloud (e.g., Padoan et al
1997).

For a completely smooth cloud, the $\sigma$-A$_V$ relation is flat with the
dispersion in $\sigma$-A$_V$ (as well as its mean value) solely determined by the
observational uncertainties in the photometry and the adopted intrinsic color of
background stars.  This dispersion will be larger in the presence of random, angular
fluctuations in extinction that are unresolved in a single map pixel.  In addition
the $\sigma$-A$_V$ relation will display a positive slope in the presence of either
unresolved, systematic gradients in extinction (Lada et al.  1999) or random, angular
fluctuations in extinction whose scale or amplitude is a function of extinction
(e.g., Padoan et al 1997).  In the case where systematic density gradients are
present in the cloud and are unresolved within an individual map pixel, the magnitude
of the slope of the $\sigma$-A$_V$ relation will depend on both the magnitude of the
density gradient and the pixel size used to construct the map (i.e., $\sigma$ $\sim$
${dA_V \over dr}\Delta r$; Lada et al.  1999).  In the case where unresolved random
spatial fluctuations are present, the magnitude of the slope will depend on the
functional dependence of the amplitude and/or angular scale of fluctuations on column
density.

We determined the $\sigma$-A$_V$ relation for Coalsack G2 at varying angular
resolutions (45, 30, 20 \& 10 arc sec) from maps made with square sampling functions
and compared the results.  We found the $\sigma$-A$_V$ relation to have a positive
slope, independent of resolution, suggesting the presence of unresolved structure on
all three spatial scales.  However, the slope was found to be identical for the 45,
30, 20 arc sec pixels with a measured value of 0.06 $\pm$ 0.003.  This suggests both
that the large scale density gradient is resolved (i.e., ${dA_V \over dr}\Delta r$
$<<$ $\sigma_{obs}$) and that the bulk of the unresolved spatial structure is random
in nature and originates on angular scales smaller than 20 arc sec.  At 10 arc sec
resolution, the slope of the $\sigma$-A$_V$ relation was found to be slightly but not
significantly flatter, 0.05 $\pm$ 0.01.  However it is possible that the small scale
structure is beginning to be resolved at this scale suggesting that random spatial
column density fluctuations have a characteristic scale of $\sim$ 10 arc sec or
smaller.  Such features appear to be discernible in the spatial map shown earlier in
figure~\ref{extmap}.

Figure~\ref{sigma-av} shows the $\sigma_{{\rm A}_V}$ -- {\rm A}$_V$ relation for
Coalsack G2 derived from our data using 20 arc sec square pixels.  As noted above,
the distribution of measurements in this diagram appears relatively flat, suggesting
that our observations have resolved much of the structure in the cloud.  However, the
existence of a measurable slope does suggest the presence of additional unresolved
structure within our 20 arc sec pixels.  For the Coalsack G2 we find a limit on the
amplitude of random, small scale ($<$ 20 arc sec) angular fluctuations in the
extinction to be $\delta{\rm A}_V /{\rm A}_V \leq $ 17\% at an A$_V$ of 15
magnitudes.  The only other cloud for which infrared extinction measurements exist
with similarly high angular and spatial resolution is B68.  For this cloud Alves et
al.  (2004) find the $\sigma$-A$_V$ relation to be essentially flat at 10 arc sec
pixel resolution and placed a limit of $\delta{\rm A}_V /{\rm A}_V < $ 5\% at an
A$_V$ of 30 magnitudes.  This suggests that structure of Coalsack G2 is not as
spatially smooth as that of the B68 cloud consistent with the impression derived from
the spatial maps of both globules as mentioned earlier.

\section{Discussion}

\subsection{The Nature of the Globule}

Our infrared extinction mapping shows that the structure of Tapia's Globule 2 in the
Coalsack cloud clearly differs in its central regions from other well studied globules and
cloud cores.  It is not as centrally condensed as B68 (ALL01), B 335 (Harvey et al.  2002)
or L 694-2 (Harvey et al.  2003).  This property of the globule was evident in the
earliest studies of its extinction profile (Jones et al.  1984).  Moreover, our
observations reveal a deep central depression in the column density profile of the cloud.
Rather than being centrally condensed, the regions of highest extinction form a clear ring
structure.  The ring cannot be the result of a change in dust properties at high
extinction, since our JHK color-color diagram shows that the extinction law is independent
of extinction to the highest observed extinctions (see also Racca et al.  2002).  Thus the
ring represents a true enhancement in the dust column density, and likely an enhancement
of the total column density since there is no reason to suspect that the gas-to-dust ratio
varies in this object.  The relation of this column density ring to the physical structure
and nature of the globule, however, depends on the geometry of the globule.

\subsubsection{Flattened Geometry and Magnetic Fields}

One possible interpretation of the observations is that the globule is a flattened, perhaps
oblate structure.  Flattened clouds are more susceptable to fragmentation than spherical
clouds and in special circumstances can form gravitationally unstable rings (e.g., Bastien
1983; Li 2001) which could fragment to form multiple star systems (Li \& Nakamura 2002).
These rings are not static structures and are found in the non-magnetic case to move inward
at the speed of sound (e.g., Bastien 1983) thus yielding a relatively short evolutionary
timescale ($\approx$ 10$^5$ yrs) for the central regions of the globule. 
However, ring formation requires very
specialized initial conditions which include a highly flattend initial state, no rotation,
a uniform or flat initial surface density gradient and a mass greatly exceeding the Jean's
mass (Bastien 1983, Li 2001).  These conditions, particularly the last one, would appear to
rule out such models for the G2 cloud.  Flattened, oblate clouds are also the expected
result of the more generalized and gradual evolution and collapse of magnetized clouds
(e.g., Mouschovias 1976).  In such an instance the cloud will be flattened along the field
lines.  For the case of Coalsack G2, the observed symmetry of the ring structure would
suggest that we are observing the ring very close to a face-on orientation with an
inclination close to 90$^o$.  Thus the magnetic field lines would be expected to be mostly
aligned along the line-of-sight, out of, or into, the plane of the sky.  However,
measurements of the infrared polarization of some 38 background stars observed through the
globule by Jones, Hyland and Bailey (1984) suggest a relatively ordered field with a
significant component of the
field lines oriented in the plane of the sky.  However, since such polarization
measurements are only sensitive to the component of polarization in the plane of the sky,
the possibility of a dominant line-of-sight component of the field still exists.
Although if it does exist, such a component would require a relatively strong line-of-sight
field, given the magnitudes of the measured polarizations ($\sim$ 1-2\%) for the
plane of the sky component.

Chandrasekhar and Fermi (1953) showed that for a magnetic field aligned primarily
in the plane of the sky the field strength, B, is related to the dispersion
in measured polarization angles ($\sigma_\Theta$), the line-of-sight velocity
dispersion of the gas ($\sigma_v$) and the gas density ($\rho$) as follows:

\begin{equation}
{\rm B} =  {\sigma_v \over \sigma_\Theta} \sqrt{4 \pi \rho}
\end{equation}

\noindent
For the case of Coalsack G2, we find from the observations of Jones et al. (1984) 
that $\sigma_\Theta$ = 0.66 radians. For a molecular hydrogen density of
\=n = 3 x 10$^3$ cm$^{-3}$ and $\sigma_v$ = 0.23 \kms, we derive B = 13 $\mu$~G for 
the cloud from the Chandrasekhar-Fermi equation above. We can also estimate the
total field strength by assuming equipartition of magnetic, gravitational and
kinetic energy in the usual manner:

\begin{equation}
B = \sqrt{{3 \pi \rho \over 2 ln2}} \ \Delta V_{NT}
\end{equation}

\noindent
where $\Delta V_{NT}$ is the non-thermal component of the observed linewidth.
For the parameters of the Coalsack G2 and with $\Delta V_{NT}$ = .53 \kms we find
B = 14 $\mu$~G. The close correspondence between the two estimates of the
field strength supports the notion that the field lies primarily in the 
plane of the sky. This would seem to argue against a highly flattened configuration
for the geometry of Coalsack G2. For the purposes of further discussion 
we therefore assume that the globule is more spherical than flattened 
in shape. 

\subsubsection{Spherical Geometry, A Core in Transition}

For a more spherical cloud geometry, the ring could be interpreted as a high density
shell.  For an isothermal or polytropic equation of state, the high density shell
would also be the location of maximum pressure in the cloud.  Such a configuration
cannot be in equilibrium with pressure balancing self-gravity, as would be the
case for the more centrally condensed Bonnor-Ebert sphere.  Thus, the shell must
be either collapsing or expanding.  An expanding shell might be expected in the
case where an additional source of internal pressure was present in the center of
the cloud.  A newly formed star with a stellar wind would do the trick, however
all of the stars we detected in the central regions of the globule appear to be
reddened background stars, none exhibit near-infrared excess or are mid-infrared
IRAS or MSX sources.

In the absence of such an additional source of internal pressure, the shell would be
expected to be in a state of inward collapse, since the weight of the cloud beyond the
ring would likely confine any outward expansion.  The close correspondence of the
observed column density profile of the globule with that of a stable Bonnor-Ebert sphere
in all but the central regions suggests that self-gravity and pressure could be in
balance at the outer edge of the ring.  The steep drop in the density profile of the
cloud in the inner regions, compared to that expected for a stable Bonnor-Ebert
configuration (figure \ref{rprofile}), indicates that the central pressure is not
sufficiently high to balance the weight of the cloud and suggests that the inner high
density ring should collapse inwards.  The timescale for such collapse is relatively
short.  The sound crossing time for the central regions is given by $\tau_a$ = r$_r$/a
where a is the speed of sound and r$_r$ the radius of the ring.  For T$_K$ = 10 K, r$_r$
= 50 arc sec, and D = 150 pc, $\tau_a$ = 2 x 10${^5}$ yrs.  The width of the C$^{18}$O
line profile is found to be 0.55 km s$^{-1}$ which is significantly in excess of that
(0.12 km s$^{-1}$ ) expected for thermal broadening in a 10 K gas.  The dynamical time
in the central regions is given by $\tau_d$ = r$_r$/$\sigma_{NT}$ = 1.5 x 10$^5$ yrs,
slightly less than the sound crossing time.  Here $\sigma_{NT}$ is the one-dimensional,
non-thermal velocity dispersion assuming a 10 K gas.  Both these timescales are less
than the estimated lifetime (10$^6$ years) of a starless core (Lee \& Myers 1999).

Another possibility is that the globule is in a state of overall oscillation.
For clouds that are in a stable equilibrium state there is a natural tendency to
oscillate upon being perturbed.  For example, Clarke and Pringle (1997) have shown that
such oscillations can arise from imbalances in the local heating and cooling rates in
cloud cores.  Moreover, more recently Matsumoto and Hanawa (2003) have found that the
presence of small amounts of rotation can cause initially unstable and collapsing cloud
cores to stablize and oscillate.  Indeed, transient central ring structures appear in
some of the oscillating clouds in their numerical simulations.  Molecular-line
observations of B68 have also suggested that non-radial surface pulsations are present
in that near- critical cloud (Lada et al 2003).  If the ring in G2 were the result of
such oscillatory behavior, the oscillations would likely move at the acoustic or sound
speed and result in time scales for significant structural change in the central regions
of the cloud comparable to that derived above for the implosion of the ring.
Although the central ring structure in G2 is out of equilibrium with its surroundings,
it is not possible from these data and existing theory to determine whether or not the
ring is unstable and imploding or oscillating around some temporary equilibrium state.
Nonetheless, these considerations do suggest that the globule is quite young and that
its presently observed structure is transient.  

Although as a whole the globule may be best described by a globally spherical geometry,
the central column density ring is probably not a spherically symmetric structure.
Models we have constructed to explain the ring as a limb-brightened spherical shell
cannot reproduce the relatively high contrast in extinction between the ring and the
central depression.  This suggests that the ring may be more toroidal than spherical in
shape.  Nonetheless, the ring would still likely be a transient structure since material
within it would not be in pressure equilibrium with gas in the central regions of the
cloud.  Independent of the exact details of the cloud geometry, it appears that the
globule is in a state of transition, perhaps in the process of transforming into a more
centrally condensed state, similar to that of typical Bok globules and dense cores that
form stars.  
%We explore this possibility further in the next section.

\subsection{Turbulence and the Origin of the Ring}

%Additional support for such a possibility may derive from consideration of our \ceio
%observations.  
The observed \ceio line toward the center of Coalsack G2 displays a
complex profile consisting of two blended components (i.e., the NLC and BLC).  The
linewidth of each blended component is super thermal indicating that significant
non-thermal motions characterize the velocity field of this cloud.  The origin of the
non-thermal component of the linewidths in G2 is likely turbulent motions in the cloud.
The three-dimensional turbulent velocity dispersion (0.15 \kms) of the NLC is slightly
more than a factor of two lower than that (0.39 \kms) of the BLC, indicating a lower
degree of turbulent motions.  The turbulent motions in the BLC are supersonic at about
Mach 2 while the turbulence in the gas emitting the narrow line is subsonic.  The NLC
also appears to be less extended than the broad component, yet it is 50\% brighter.

The narrow-line feature appears to be absent in observations made with lower angular
resolution, in particular, it is not present in H$_2$CO absorption measurements obtained
with a beamwidth of 4.4 arc min, an order of magnitude larger than that of our CO
observations (Brooks et al.  1976).  This suggests that the NLC has been heavily beam
diluted in the H$_2$CO observations and thus arises in a region less than, or on the
order of, 1-2 arc min in extent.  In this region the supersonic turbulent motions that
characterize the bulk of the cloud have been largely dissipated.  Since:  1) the
inferred scale of the narrow line region is comparable to the size of the high column
density ring, and 2) the coordinates of the CO observation correspond to a location
centered on the southwest corner of the high extinction ring, we suggest that the NLC
originates within the ring.  A high resolution ( $\sim$ 30 arc sec) map of \ceio
emission would enable a direct test of this particular hypothesis.  If, however, this
proposition is correct, then it is possible that the ring is a high density structure
that formed by the collision and compression of turbulent gas elements within the
globule.  This process likely resulted in shock compression of the gas and the
consequent dissipation of turbulent energy which produced the subsonic velocity
dispersion in the NLC.  In this picture, the dissipation of turbulence results in the
formation of the ring structure and the production dense gas in the central regions of
the globule.  Presumably, as more turbulence is dissipated in the cloud, the cloud would
adjust to a more centrally concentrated state.  This turbulent model for the
origin of the ring would also be consistent with the non-spherical geometry of the ring,
including the fact that the NLC and BLC differ in velocity, something that would not be
expected in spherical symmetry.  This interpretation of the motions of the molecular gas
provides additional support for the notion that the G2 core is in an early state of
evolutionary development.

\section{Summary and Concluding Remarks}

A number of considerations from the analysis of our observations of Tapia's Globule~2 in
the Coalsack suggest that the globule is a bound core in a state very close to overall
dynamical equilibrium.  Perhaps the most compelling of these is the finding that the
measured gas motions in the globule, derived from the linewidth of \ceio emission,
indicate that the cloud is gravitationally bound and may even be near virial
equilibrium.  Moreover, the density gradient of the cloud, beyond the central ring, is
consistent with that of a sub-critical BE cloud in which thermal and non-thermal
pressure balance self-gravity.  This, in turn, is consistent with the fact that the
cloud's mass is found to be less than the Jean's mass.  In addition the globule's
structure is well resolved by our extinction observations and appears to be relatively
smooth, though less smooth than B 68, the only other cloud studied in such detail by
infrared extinction measurements.

However, the G2 core differs from typical bound cloud cores in some significant ways.
The peak extinction (A$_V$ $\approx$ 12 mag.)  is relatively low for a dense core.  Most
importantly, the density structure of the central regions of the G2 cloud departs
significantly from that of a typical cloud core and from the predictions for a centrally
condensed BE configuration in hydrostatic equilibrium.  In particular, the observed
column density distribution is characterized by a deep central depression bordered by a
well-defined ring of high column density and high pressure.  The central ring is likely
out of equilibrium with the gas in the very center of the cloud and consequently appears
to be a transient feature with an evolutionary time scale of order 10$^5$ years which is
less than the expected million year lifetime of a starless core.  This would seem to
indicate that the central regions of the cloud are in a phase of structural transition
and evolving to a new physical configuration.  Given that the central column density
depression represents a mass deficit between the observed and best-fit BE density
profile of only about 1-2\% the total cloud mass, it seems plausible to assume that the
cloud will evolve to a more relaxed, BE-like, configuration. The gradual dissipation
of turbulence in the central regions of the cloud may be driving this transition.
Thus, G2 appears to be a cloud core in an early stage of formation.

The Coalsack complex has long been known to be unusual among molecular clouds in its
lack of any significant star forming activity (Nyman et al.  1989).  Earlier CO
observations have shown that it is also unusual in its relative paucity of dense gas
(Kato et al.  1999).  These facts are clearly not unrelated since surveys of molecular
clouds have demonstrated that stars form exclusively in dense gas (e.g., Lada 1992).
The Coalsack complex is plausibly a very young molecular complex.  The fact that we find
the structure of the most opaque globule in the Coalsack cloud to clearly differ from
the steep centrally condensed configurations of star forming cores and globules in other
clouds, appears consistent with the idea that this core is also quite young.  Exactly
how early this core is in its development is difficult to determine.  However, many
starless cores, for example, B68 (ALL01) and L694-2 (Harvey et al 2004), are known to be
characterized by a high degree of central concentration, often similar to star forming
cores.  If such structure is more typical of starless cores then we would suspect that
G2 represents a rare example of an extremely early phase of dense core evolution.
Investigation of the detailed structures of a larger sample of starless cores is clearly
needed to better quantify this conjecture.  However, we note that although starless
cores in molecular clouds are relatively common (Lee \& Myers 1999), cloud complexes
without star formation are extremely rare.  Together these factors suggest the extreme
youth of the Coalsack complex and indicate that this region constitutes
an important laboratory for future studies of the poorly understood processes of cloud
and core formation.

\acknowledgments

We thank Karoline P.  Pershell for able assistance with the infrared data reduction.  We
acknowledge Andi Burkert, Richard Klein, Alyssa Goodman, Frank Shu and Zhi-Yun Li for
useful discussions and an anonymous referee for comments which improved the paper.  This
work was supported by NASA Origins Grants NAG-9520 and NAG-13041 and by a UT Martin Faculty
Research Grant.

\clearpage


\begin{figure}
%\plotone{f1.eps}
%\vskip -2.0in
{\bf Full resolution figures and text can be downloaded from \\
\url{http://cfa-www.harvard.edu/$\sim$clada/preprints.html}}
\vskip 1.3in

\caption{DSS optical image of Globule 2, the most opaque core 
in the Coalsack complex. The box outlines the 
region surveyed in the infrared with the NTT. 
\label{optical}}
\end{figure}

\clearpage 

\begin{figure}
%\centerline{\epsfig{figure=f3.eps,width=4in,angle=90}}
%\plotone{f3.eps}
%\vskip -1.0in
{\bf Full resolution figures and text can be downloaded from \\
\url{http://cfa-www.harvard.edu/$\sim$clada/preprints.html}}
\vskip 1.3in

\caption{Azimuthally averaged radial extinction (column density) profile
for Coalsack Globule 2. Also plotted is the expected column density 
profile of the Bonnor-Ebert sphere that best fits the observations
between radii of 55 - 400 arc sec.
\label{rprofile}}
\end{figure}

\clearpage 

\clearpage 

\begin{figure}
%\centerline{\epsfig{figure=f5.eps,width=4in,angle=90}}
%\plotone{f5.eps}
%\vskip -1.0in
{\bf Full resolution figures and text can be downloaded from \\
\url{http://cfa-www.harvard.edu/$\sim$clada/preprints.html}}
\vskip 1.3in
\caption{The $\sigma$ - A$_V$ relation for Coalsack Globule 2 derived from
measurements obtained with a square spatial sampling function 20 arc sec wide.
The relation is relatively flat indicating that the structure in the cloud
has been mostly resolved at this resolution and that the cloud is relatively
smooth on this angular scale. 
\label{sigma-av}}
\end{figure}

\clearpage

%% If you are not including electronic art with your submission, you may
%% mark up your captions using the \figcaption command. See the 
%% No more than seven \figcaption commands are allowed per page, 
%% so if you have more than seven captions, insert a \clearpage 

%% The following command ends your manuscript. LaTeX will ignore any text

\end{document}